\documentclass{article} 
\usepackage{jcappub}

\usepackage{epstopdf}
\usepackage{graphicx}
\usepackage{color}
\usepackage{hyperref}

\newcommand{\be}{\begin{equation}}
\newcommand{\ee}{\end{equation}}
\newcommand{\bea}{\begin{eqnarray}}
\newcommand{\eea}{\end{eqnarray}}
\newcommand{\bdm}{\begin{displaymath}}
\newcommand{\edm}{\end{displaymath}}
\newcommand{\de}{d}

\newcommand{\Lam}{\Lambda}
\newcommand{\sig}{\sigma}
\newcommand{\Om}{\Omega}

\newcommand{\dc}{\delta_{cr}}

\def\Ms{\, h^{-1} \, M_\odot}
\def\Mpc{\, h^{-1} \, {\rm Mpc}}

\def\cGpc{\, h^{-3} \, {\rm Gpc}^3}
\def\kMpc{\, h \, {\rm Mpc}^{-1}}

\def\sigm{\sigma_{mm}}
\def\sigc{\sigma_{cc}}
\def\sigem{\sigma_{8,mm}}
\def\sigec{\sigma_{8,cc}}

\title{Cosmology with massive neutrinos II: on the universality of the halo mass function and bias}

\author[a]{Emanuele Castorina,}
\author[b,c]{Emiliano Sefusatti,}
\author[c,d]{Ravi K. Sheth}
\author[e]{Francisco Villaescusa-Navarro,} 
\author[e,f]{and Matteo Viel}

\affiliation[a]{SISSA- International School for Advanced Studies,
Via Bonomea 265, I-34136 Trieste -- Italy}
\affiliation[b]{INAF - Osservatorio Astronomico di Brera, via E. Bianchi 46, I-23807 Merate (LC) -- Italy}
\affiliation[c]{The Abdus Salam International Center for Theoretical Physics, strada costiera 11, I-34151 Trieste -- Italy}
\affiliation[d]{Center for Particle Cosmology, University of Pennsylvania, 209 S. 33rd St., Philadelphia (PA) 19104 -- USA}
\affiliation[e]{INAF - Osservatorio Astronomico di Trieste, via Tiepolo 11, I-34143 Trieste -- Italy}
\affiliation[f]{INFN - sezione di Trieste, via Valerio 2, I-34127 Trieste -- Italy}

\emailAdd{ecastori@sissa.it}
\emailAdd{emiliano.sefusatti@brera.inaf.it}
\emailAdd{sheth@ictp.it}
\emailAdd{villaescusa@oats.inaf.it}
\emailAdd{viel@oats.inaf.it}

\abstract{We use a large suite of N-body simulations to study departures from universality in halo abundances and clustering in cosmologies with non-vanishing neutrino masses.  To this end, we study how the halo mass function and halo bias factors depend on the scaling variable $\sigma^2(M,z)$, the variance of the initial matter fluctuation field, rather than on halo mass $M$ and redshift $z$ themselves.  We show that using the variance of the cold dark matter rather than the total mass field, i.e., $\sigma^2_{cdm}(M,z)$ rather than $\sigma^2_{m}(M,z)$, yields more universal results.  
Analysis of halo bias yields similar conclusions:  When large-scale halo bias is defined with respect to the cold dark matter power spectrum, the result is both more universal, and less scale- or $k$-dependent.  These results are used extensively in Papers I and III of this series.}


\begin{document}
\maketitle

\section{Introduction}

The abundance by mass of galaxy clusters, and of the dark matter halos which surround them, is a major tool for cosmological parameter estimation (see \cite{Borgani2008} for a review).  Large catalogs are now available from optical \cite{RozoEtal2013}, X-ray \cite{MehrtensEtal2012,WillisEtal2013} and Sunyaev-Zel'Dovich (SZ) observations \cite{HasselfieldEtal2013, ReichardtEtal2013, Planck2013SZ}. Previous work has shown that, in neutrino-less $\Lambda$CDM models, the halo mass function over a wide range of redshifts and background cosmologies can be scaled to an almost universal form \cite{PressSchechter1974, ShethTormen1999}.
This universality is particularly useful, as it vastly simplifies analyses of observed datasets.  However, it has been known for some time that this sort of universality should only be an approximation \cite{ShethTormen1999, ParanjapeEtal2013}, and departures from universality of about the expected level have indeed been detected in recent simulations \cite{TinkerEtal2008, CrocceEtal2010}.  In this work we study the issue of universality, and departures from it, in the context of neutrino cosmologies.  
While these are interesting in their own right -- see Paper I of this series \cite{Villaescusa-NavarroEtal2013B} for a more detailed introduction to neutrino physics and its cosmological implications -- our study is motivated in part by the tension reported by the Planck collaboration between their temperature and cluster count measurements \cite{Planck2013SZ}.  Paper III of this series \cite{CostanziEtal2013B} is dedicated to the implications of our findings for such analyses.  

The shape of the galaxy power spectrum and correlation function are also sensitive to the underlying cosmology, and can be used to put strong constraints on cosmological parameters \cite{AndersonEtal2012, SanchezEtal2012}. In particular, such observables are able to provide upper bounds to the sum of neutrino masses, \cite{Hannestad2003, ReidEtal2010, ThomasAbdallaLahav2010, SwansonPercivalLahav2010, SaitoTakadaTaruya2011, DePutterEtal2012B, XiaEtal2012, RiemerSorensenEtal2012, GiusarmaEtal2013, ZhaoEtal2013}.  In the Halo Model of large scale structure \cite{CooraySheth2002} these are a consequence of the fact that the spatial clustering of dark matter halos is biased with respect to that of the total mass, and the details of how this bias depends on halo mass depend on the background cosmological model.  Therefore, a related goal of this work is to provide an analysis of halo bias in neutrino cosmologies.  

This paper is organized as follows. Section~\ref{sec:info} summarizes the role of neutrinos in cosmological halo formation, and argues that universality should be more apparent in the cold dark matter component than in the total.  Section~\ref{sec:sim} describes the simulations and the halo catalogs employed in this work.  Section~\ref{sec:massf} presents our measurements of the Friends-of-Friends halo mass function and discusses its dependence on neutrino mass.  Section~\ref{sec:bias} presents a similar analysis of halo bias.  We conclude, in Section~\ref{sec:conclusions}, that halo abundances and clustering are indeed more universal if one works with the cold dark matter component only, and that failure to account for this will lead to inaccurate constraints on neutrino cosmologies.

\section{Background}\label{sec:info}

Through-out this work we will use ``cold'' to denote the actual cold dark matter component as well as baryons.  We distinguish this CDM component from neutrinos which are characterized by large thermal velocities.

\subsection{Cosmological neutrinos}
\label{sec:nus}

For excellent reviews of neutrinos in cosmology see \cite{LesgourguesPastor2006, LesgourguesEtal2013}.  For our purposes, the key elements are as follows.  

Neutrinos decouple in the early universe as ultra-relativistic particles.  They then behave like radiation, until they become non-relativistic at a redshift $z_{nr}$ given by  
\be
 1+z_{nr}(m_{\nu}) \simeq 1890 \left(\frac{m_{\nu}}{1\,{\rm eV}}\right)\,,
\ee
where $m_{\nu}$ is the neutrino mass.  Thereafter, the total dark matter (DM) background density is given by $\Om_m = \Om_{cdm} + \Om_{\nu}$.  It is convenient to think of a perturbation $\delta_m$ in the total DM field as a weighted sum of the fluctuations $\delta_{cdm}$ and $\delta_\nu$ in the CDM and $\nu$ fields:  
\be
\label{eq:OmDM}
 \delta_m \equiv (1-f_{\nu})\, \delta_{cdm} + f_{\nu}\, \delta_{\nu}\,,\quad
 {\rm where}\quad f_{\nu} \equiv  \Om_{\nu}/\Om_m.
\ee
In what follows, we will use $P_{mm}(k)$ to denote the power spectrum of the total field, $P_{cc}(k)$ and $P_{\nu\nu}(k)$ the power spectra of the CDM and $\nu$ fields, and $P_{\nu c}(k)$ the cross-power between the two fields.  Therefore, 
\be 
 P_{mm}(k) = (1-f_{\nu})^2\,P_{cc}(k) + f_{\nu}^2 P_{\nu\nu}(k) + 2f_{\nu}(1-f_{\nu})P_{\nu c}(k).
\ee

The growth of neutrino fluctuations is governed by their free streaming length $\lambda_{fs}$, below which perturbations are washed out. At early times ($z>z_{nr}$) $\lambda_{fs}$ is of the order of the horizon scale.  However, after the non-relativistic transition, 
\be
 \lambda_{fs} (m_\nu,z) = a \left(\frac{2 \pi }{ k_{fs}}\right)
   \simeq 7.7 \frac{1+z}{(\Om_{\Lam}+\Om_m(1+z)^3)^{1/2}}
         \left(\frac{1\,eV}{m_{\nu}}\right)\, \Mpc\,.
\ee
The free streaming length has a minimum at $z=z_{nr}$
\be
 k_{nr} = k_{fs} (z_{nr})
      \simeq 0.018 \Omega_m^{1/2} \left(\frac{m_{\nu}}{1\,eV}\right) \kMpc\,.
\ee
At $k\ll k_{nr}$, the CDM and $\nu$ fields are tightly coupled, so $P_{cc} \approx P_{\nu\nu} \approx P_{\nu c}$ making $P_{mm}\to P_{cc}$.  At sufficiently large $k$, there is no power in the $\nu$ field, so $P_{mm}\to (1-f_{\nu})^2\,P_{cc}$.  Thus,
\be\label{eq:Pcdm}
 P_{mm}(k) = \begin{cases}
           P_{cc}(k) & \text{if } k<k_{nr} \\
           (1-f_{\nu})^2\,P_{cc}(k) & \text{if } k\gg k_{nr}\,.
          \end{cases}
\ee
These limiting cases suggest that the CDM and matter fields are actually rather similar when $f_\nu\ll 1$ is small.  One measure of this is the cross-correlation coefficient 
\be
 r_{cm} \equiv \frac{P_{cm}}{\sqrt{P_{cc}P_{mm}}}.
\ee
Since 
\be
P_{cm}=\langle \delta_c \delta_m \rangle = \langle \delta_c ( (1-f_{\nu}) \delta_c + f_{\nu} \delta_{\nu} )\rangle =  (1-f_{\nu}) P_{cc} + f_{\nu} P_{c\nu}\,,
\ee
and 
\bea
 \sqrt{P_{cc}\,P_{mm}} & = & \sqrt{(1-f_{\nu})^2 P_{cc}^2 + f_{\nu}^2 P_{\nu \nu}P_{cc} + 2f_{\nu}(1-f_{\nu})P_{c\nu} P_{cc}}\nonumber\\
 & = &  (1-f_{\nu}) P_{cc} + f_{\nu} P_{c\nu} + {\mathcal O}(f_{\nu}^2)
   =  P_{cm} + {\mathcal O}(f_{\nu}^2)\,,
\eea
differences from $r_{cm}\simeq 1$ only appear at second order in $f_{\nu}$.  

Finally, note that the growth rate of cold dark matter (CDM) inhomogeneities is slower in massive neutrino cosmologies; during the matter dominated era \cite{BondEfstathiouSilk1980}
\be\label{eq:growth}
 \delta_{cdm} \propto a^{1-\frac{3}{5} f_{\nu} }\quad {\rm for}\quad k > k_{nr}. 
\ee
If the total matter density in eq.~(\ref{eq:OmDM}) is fixed, then the total matter power spectrum in a massive neutrino model, $P_{mm}(k)^{f_\nu}$ is reduced by a constant factor on scales $k \gg k_{nr}$ and small values of $f_{\nu}$ \cite{HuEisensteinTegmark1998, LesgourguesPastor2006}
\be
 \frac{P_{mm}(k;f_\nu)}{P_{mm}(k;f_\nu = 0)} \simeq 1-8 f_{\nu}\,.
\ee

\subsection{Universality in the CDM-component}
\label{sec:uni}

At any redshift $z$ the comoving number density of halos per unit mass, $n(M)$, can be written in the following form
\be\label{eq:f}
 n(M) =  \frac{\rho}{M}\,f(\sigma,z) \frac{\de \ln\sigma^{-1}}{\de M} \,, 
\ee
where
\be\label{eq:sigma}
 \sigma^2(M,z) = \int d^3k\, P(k,z)\,W_R^2(k)\,
\ee
is the r.m.s.~of the linear density field smoothed on a scale $R$ with a filter function $W(kR)$, $\rho$ is the comoving background density. The relation between the smoothing scale $R$ and the halo mass $M$ is dictated by the choice of the filter function, being given by
\be
 M \equiv \rho \int \de^3 x \,W(x,R)\,.
\ee
In this context, we will define the mass function to be universal when $f(\sigma,z) = f(\sigma)$, i.e. the function $f(\sigma)$ does not depend on redshift.
This shows that the quantity which is expected to be nearly universal is not $n(M)$ itself but 
\be\label{eq:vfv}
 \nu\,f(\nu) \equiv \frac{M^2}{\rho}\,n(M)\,\frac{\de\ln M}{\de \ln \nu},
\ee
where $\nu\equiv \dc/\sigma$, for some constant $\dc$ which we will discuss shortly.  (It is unfortunate that this scaling variable is called $\nu$ when it has, of course, nothing to do with neutrinos!  We trust this will not lead to confusion.)  Operationally, one measures this quantity by first transforming $M$ to $\ln\nu$, and then binning the counts in $\nu$ upon giving each halo a weight which equals $M/\rho$.  

Without previous knowledge of the effects that a non-vanishing neutrino mass could have on the process of halo formation it is not obvious what to use in eq.~(\ref{eq:f}) for the quantities $\rho, M$ and $\sigma$, since they can be defined either in terms of all dark matter species or in terms of the cold one alone.  All we will justify later, we identify halos in simulations by using the CDM component only. In this case, one would define the relation between CDM halo mass and the scale of a TopHat filter by 
\be\label{eq:M-R}
 M = \frac{4\pi}{3}\,\rho_{cdm}\,R^3 \,.
\ee
Then, since the $M$ in eq.~(\ref{eq:f}) is really $M_{cdm}$, the $\rho$ in eq.~(\ref{eq:f}) should be replaced by $\rho_{cdm}$, i.e. the {\em cold} dark matter density. This choice is consistent with previous work \cite{BrandbygeEtal2010, MarulliEtal2011, Villaescusa-NavarroEtal2013}, where it is shown that the halo counts in $f_\nu\ne 0$ simulations are in better agreement with known (i.e. $f_\nu=0$ based) fitting formulae if $\rho_{cdm}$ is used.  

The last piece of information we need is the appropriate quantity to use for $\sigma$ in eq.~(\ref{eq:f}).  It is tempting to assume that the relevant quantity is $\sigma_{cc}$, which should be computed by setting $P=P_{cc}$ in eq.(\ref{eq:sigma}).  We believe this is well-motivated because the scales associated with halo formation are typically $\gg k_{nr}$, so it is reasonable to treat the CDM as though it alone is clustering in an effective background cosmology which depends on the large scale value of $\rho_\nu$.  Moreover, studies of spherical halo collapse suggest that what really matters for halo formation is the ratio $\dc/\sigma$, where $\dc$ is the density which linear theory predicts is associated with nonlinear halo collapse \cite{PressSchechter1974, BondEtal1991, ShethTormen1999}.  When $f_\nu = 0$ then $\dc\approx 1.686$ is only a very weak function of $(\Omega,\Lambda)$ (e.g. \cite{KitayamaSuto1996}), so one might expect the dependence of $\dc$ on $\Omega_{eff}(f_\nu)$ and $\Lambda_{eff}(f_\nu)$, and hence on $f_\nu$ itself to be negligible.  \cite{IchikiTakada2012} have confirmed that the effects of massive neutrinos on $\dc$ are less than $1\%$ for the range of $f_\nu$ we will consider.  That is to say, in these massive neutrino models, the physically relevant quantity $\dc/\sigma$ is really $\dc/\sigma_{cc}$, and since $\dc$ is almost independent of $m_\nu$, the scaling variable is actually just $\sigma_{cc}$.  

We emphasize that if neutrino perturbations had played a role in the collapse of regions, as happens for warm DM or clustering quintessence, then we would not have been allowed to simply replace $\sigma$ with $\sigc$.  In these other cases, a more complicated analysis (following methods outlined in \cite{CastorinaSheth2013}) would be needed.  Note that a model which uses $\sigc$ predicts more halos than one with $\sigm$, since (after the non-relativistic transition of neutrinos) $\sigc \ge \sigm$ for all relevant smoothing scales (see eq.~\ref{eq:Pcdm}).

\section{Simulations}
\label{sec:sim}

\subsection{Cosmological models and N-body runs}

The N-body simulations used in this paper were run using a modified version of the \texttt{GADGET-3} code, described in \cite{VielHaehneltSpringel2010}.  In this code, neutrinos are treated exactly as cold dark matter particles, but they are assigned, in the initial conditions, large thermal velocities drawn from a Fermi-Dirac distribution. Linear transfer functions from \texttt{CAMB} \cite{LewisChallinorLasenby2000} are used to generate initial conditions at $z=99$ using the Zel'dovich approximation \cite{Zeldovich1970}.  In practice, we use a modified version of \texttt{NGenIC} that gives to neutrinos the same random phases as for the cold dark matter:  I.e. we assume adiabatic initial conditions.  The large initial redshift means that transient effects on the halo mass function and bias should be negligible \cite{Scoccimarro1998, CroccePueblasScoccimarro2006}.
We present results for cosmologies with three different values of neutrino masses: $\sum m_{\nu}=0$, $0.3$ and $0.6$ eV; from now on, we will use $m_\nu$ to mean $\sum m_{\nu}$. 
     
A first set of simulations (Set A) shares the following cosmological parameters: $\Omega_b = 0.05$, $\Omega_{\Lambda}=0.7292$, $h=0.7$, $n_s =1 $ and $A_s = 2.43 \times 10^{-9}$, with zero curvature. The total matter density is also fixed to $\Omega_m = \Omega_b + \Omega_{c} + \Omega_{\nu}=\ 0.2708$, such that the cold dark matter density changes as $\Omega_\nu$ varies. In addition, the shared value for the amplitude of initial fluctuations $A_s$ results in different values for the amplitude of cold and total matter perturbations at late times, parametrized, for instance, respectively by $\sigec$ and $\sigem$. 

A second set of three simulations (Set B) explores possible degeneracies of the initial amplitude $A_s$ with the value of $m_\nu$, still keeping $\Omega_b$, $\Omega_{m}$, $h$ and $n_s =1$ at the same values. In Set B, two massless neutrinos cosmologies where we changed the initial amplitude $A_s$ to match the $\sigma_8$ of the DM component and CDM component of the $m_{\nu}=0.6$ model in Set A. The third one has $m_{\nu}=0.6$ eV but matches $\sigec$ to the massless neutrino model in Set A. Table~\ref{tab:par} summarizes the different sets of parameters used in this paper.

\begin{table}[t]
\begin{center}
\begin{tabular}[]{l | c c  c c c  c c}
&  $\sum m_{\nu}$[eV]   & $\Om_c$ & $f_{\nu}$ & $\sigem$ & $\sigec$ &$m^{c}_p [\Ms]$ &$m^{\nu}_p [\Ms]$\\                    
\noalign{\smallskip}\hline \hline \noalign{\smallskip}
\multicolumn{7}{l}{Set A}\\
\noalign{\smallskip}\hline \hline \noalign{\smallskip}
 H0       & 0.0 & $0.2208$ & $0.000$ & $0.832$ &$0.832$ & $5.60 \times 10^{11}$ & $-$\\
 H3       & 0.3 & $0.2142$ & $0.024$ & $0.752$ &$0.768$ & $5.46 \times 10^{11}$ & $1.36 \times 10^{10}$\\
 H6       & 0.6 & $0.2076$ & $0.048$ & $0.675$ &$0.701$ & $5.33 \times 10^{11}$ & $2.72 \times 10^{10}$\\
\noalign{\smallskip}\hline \hline \noalign{\smallskip}
\multicolumn{7}{l}{Set B}\\
\noalign{\smallskip}\hline\noalign{\smallskip}
 H0s8     & 0.0 & $0.2208$ & $0.000$ & $0.675$&$0.675$ & $5.60 \times 10^{11}$ & $-$\\
 H0s8-CDM & 0.0 & $0.2208$ & $0.000$ & $0.701$ &$0.701$ & $5.60 \times 10^{11}$ & $-$\\
 H6s8     & 0.6 & $0.2076$ & $0.048$ & $0.832$ &$0.864$ & $5.33 \times 10^{11}$ & $2.72 \times 10^{10}$\\
\hline \noalign{\smallskip}
\end{tabular}
\caption{\label{tab:par} Summary of cosmological parameters and derived quantities for the six models assumed for our N-body simulations. The values $\Omega_b = 0.05$, $\Omega_{m}=0.2708$, $h=0.7$, $n_s =1$ are shared by all models.}
\end{center}
\end{table}

For each model we performed eight realizations with different random seeds of a cubic box of linear size $1000$ $\Mpc$ with $512^3$ CDM particles and $512^3$ neutrino particles, so that for each model we reach a combined effective volume of $8\cGpc$. Auto and cross power spectra of the different species (CDM, DM, neutrinos) are computed at $z=2$, 1, $0.5$ and $0$, where halo catalogs are also produced.

\subsection{Halo finder}
\label{sec:FoF}

Halos in each simulation are identified by running the Friends-of-Friends (FoF) algorithm \cite{DavisEtal1985} on the CDM particles only, with linking length set to $b=0.2$ times the mean CDM-particle distance.  The FoF halo masses are corrected for the statistical noise arising from particle discreteness following \cite{WarrenEtal2006}, by setting $M_{halo} = N_{corr}\,m_p^{c}$, $m_p^{c}$ being the cold dark matter particle mass and 
\be\label{eq:warr}
 N_{corr}= N_p\, ( 1-N_p^{-0.6} ),
\ee
where $N_p$ is the original number of particles linked together than the FoF algorithm.  For the minimum number of particles per halo considered in this paper, $N_p = 32$, this correction can be larger than $15\%$.  Since eq.~(\ref{eq:warr}) is a correction to $N_p$ only, halos of the same mass in different cosmologies undergo different corrections because $m_p^{c}$ is different in the different runs (because $\Omega_{c}$ increases when $\Omega_\nu$ decreases). Halo power spectra and cross halo-matter power spectra are computed applying two different cuts in mass, $M>2\times10^{13}\Ms$ and $M>4\times10^{13}\Ms$.
The shot-noise contribution to the halo power spectra at $z=2$ is large, so we only present results for  $z=1$, $0.5$ and $0$. 
 
One might have worried that if neutrinos affect halo profiles, then eq.~(\ref{eq:warr}) should be slightly modified in neutrino cosmologies.  Simulations have indeed shown that halos in neutrino cosmologies are less concentrated than their counterparts in standard $\Lam$CDM models, because their formation time is delayed due to the suppression of the power spectrum induced by neutrinos \cite{Villaescusa-NavarroEtal2013, BrandbygeEtal2010}. However this effect is rather small and it can be safely neglected.

Another possible choice would be to run the FoF algorithm on all the particles in the box. This test is discussed in detail in Paper I \cite{Villaescusa-NavarroEtal2013B}. The halo power spectra (for the two mass thresholds defined above), differ by less than $0.5\%$, in good agreement with the expectation that CDM particles contribute the most to the mass of halo and hence to its center of mass. However some discrepancies are present in the halo mass function, especially at low masses, produced by spurious assignment of neutrinos to halos. For small halo masses, in fact, neutrinos are not bound, they free-stream due to their large velocities, but the FoF finder wrongly assigns them to halos. This contamination is more important at low neutrino number densities like those considered in this paper. A proper procedure which excludes unbound neutrino particles would therefore lead to differences in the halo power spectrum that are expected to be even smaller than those found in our simple test. For high-mass halos the contribution of both bound and unbound neutrinos to the total mass is small, typically less than $0.5\%$. In the following we will always consider FoF CDM-only halos, but see Paper I for further details and tests.

\section{Halo mass function}
\label{sec:massf}

In this section we discuss the halo mass function in massive neutrino cosmologies, comparing theoretical predictions with results from the N-body simulations described in the previous Section~\ref{sec:sim}.  The implications for the analysis of observed cluster catalogs are presented in a companion paper \cite{CostanziEtal2013B}. The results of this section will be of crucial importance for the power spectra and bias analysis of Section~\ref{sec:bias}.

\begin{figure}[t!]
\begin{center}
\includegraphics[width=0.98\textwidth]{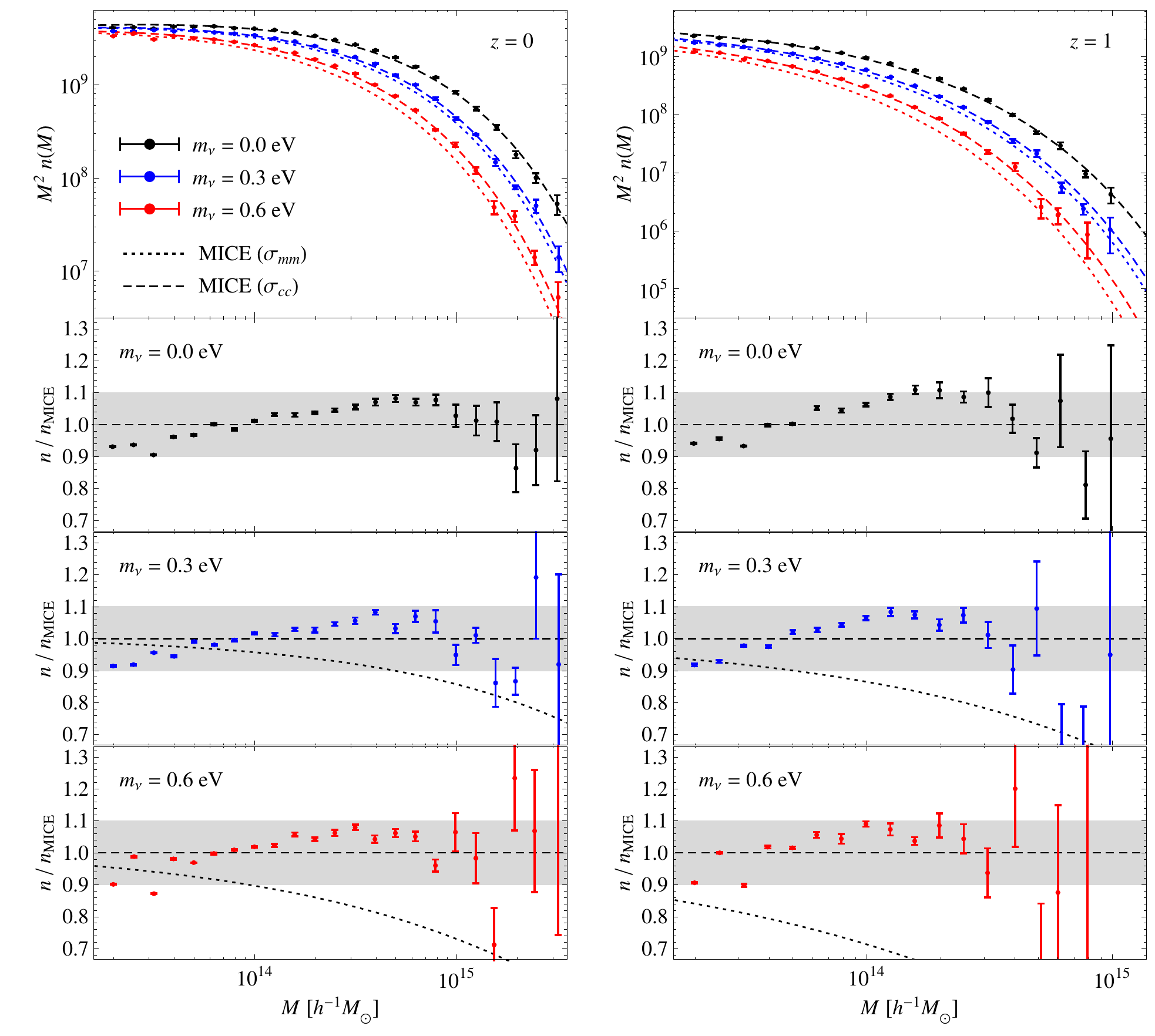}
\end{center}
\caption{\label{fig:mfM}
Halo mass function for the three models of set A at redshift $z=0$ ({\em left panels}) and $z=1$ ({\em right panels}). Top panels show the quantity $M^2\,n(M)$ as a function of mass, together with the  predictions of the MICE fitting formula \cite{CrocceEtal2010} using $\sigma=\sigm$ ({\em dotted curves}) and $\sigma=\sigc$ ({\em dashed curves}). Black, blue and red data points correspond respectively to the $m_\nu=0$, $0.3$ and $0.6$ eV results. Lower panels show the residuals of measurements with respect to the MICE formula with $\sigma=\sigc$. Here, as in the following figures, all symbols show the mean over the eight realizations while the error bars show the uncertainty on this mean.}
\end{figure}

The uppermost panels of Figure~\ref{fig:mfM} show $M^2\,n(M)$ as a function of halo mass $M$ for the three cosmologies with $m_\nu=0$, $0.3$ and $0.6$ eV (black, blue and red symbols); the left and right columns show results at $z=0$ and $z=1$.  In each panel, the symbols show the mean over the eight realizations for each cosmology and redshift, with error bars showing the uncertainty on the mean.  These are compared with the fitting formula which describes the MICE simulations:  
\be\label{eq:fMICE}
 f(\sigma,z) = A(z)\left[\sigma^{-a(z)}+b\right]\,e^{-c(z)/\sigma^2}\,,
\ee
where the parameters $A(z)$, $a(z)$, $b(z)$ and $c(z)$ depend on redshift (we use the values from \cite{CrocceEtal2010}). 

For the cosmologies with massive neutrinos, we do this in two ways, by setting $\sigma=\sigm$ (dotted curves) or $\sigc$ (dashed curves)\footnote{To partially remove finite-volume effects, we set the lower cut-off in the integral of eq.~(\ref{eq:sigma}) to the fundamental frequency of the box, $k_F=2\pi/(1,000\Mpc)$. Another possible correction for the finite size of the simulation box could be to measure the linear power spectrum from the box itself once the initial displacements were generated instead of using Boltzmann codes (see, e.g. \cite{ReedEtal2007, WatsonEtal2013}). This method has the advantage of removing cosmic variance and volume effects, but it gives a different $\sigma$-$M$ relation for each box that one has then to average over. For practical reasons we assume the linear power spectra to be given by the \texttt{CAMB} predictions for each model and we always assume the mass-scale relation as in eq.~(\ref{eq:M-R}).}. I.e., the dotted and dashed curves represent the assumptions that the relevant sigma is the rms fluctuation in total density field or the CDM component respectively. For the $m_\nu=0$ eV case, where $P_{cc}\equiv P_{mm}$, we only show a dashed curve.  The lower panels of Figure~\ref{fig:mfM} show the residuals with respect to the $\sigc$-based curve, separately for the three cosmologies of Set A, second to fourth row.  

We notice, in the first place, a small (less than 10\%) discrepancy between our $\Lambda$CDM, $m_\nu=0$ eV simulations and the MICE fit over the whole relevant mass range. As this discrepancy is about the same at higher redshifts, we conclude that our simulations show $z$-dependent departures from universality that are close to those observed in the MICE analysis.  We will return to this shortly. Agreement with the MICE predictions is not, in any event, the focus our attention.  More interesting, is that the difference with respect to the MICE fits remain roughly the same for all values of $m_\nu$ when $\sigc$ is used, whereas they grow significantly when $\sigm$ is used instead (dotted curves).

\begin{figure}[t!]
\begin{center}
\includegraphics[width=.98\textwidth]{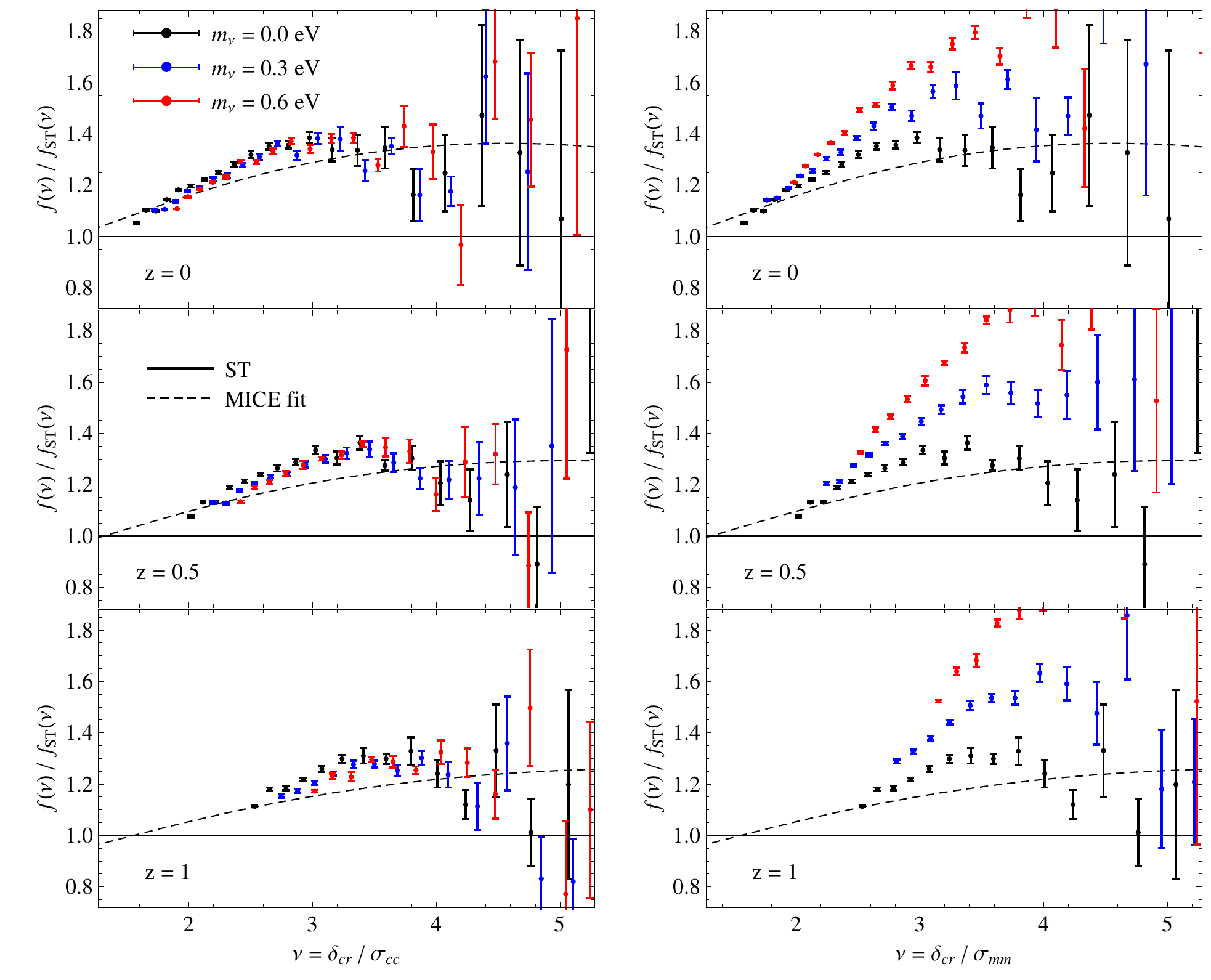}
\end{center}
\caption{\label{fig:mfnu}
Ratio of the measured $\nu\,f(\nu)$ to the ST formula. Left panels use $\nu=\dc/\sigc$ while right panels use $\nu=\dc/\sigm$. Different panels, top to bottom, show the different redshifts $z=0$, 0.5 and 1 with all neutrino masses (distinguished by color) shown together. Dashed curves show the MICE fit of \cite{CrocceEtal2010}.}
\end{figure}

To better highlight the difference between the two descriptions, it is convenient to compare measurements of the differential mass function directly as a function of the variable $\nu\equiv\dc/\sigma$, for which the mass function is given by eq.(\ref{eq:vfv}).
Such a comparison is made in Figure~\ref{fig:mfnu} which presents the same measurements of Figure~\ref{fig:mfM}, this time in terms of $\nu f(\nu)$. In particular, each panel shows the ratio of $\nu f(\nu)$ to $0.322\,[1 + (0.7\nu^2)^{-0.3}]\,\nu\sqrt{1.4/\pi}\exp(-0.7\nu^2/2)$, the universal function of Sheth \& Tormen \cite[ST,][]{ShethTormen1999}) from three simulations with different neutrino masses at a given redshift ($z=0$, $0.5$ and $1$, top to bottom).  
Also shown is the prediction from the MICE fit of \cite{CrocceEtal2010}, which explicitly depends on redshift, as a dashed black curve. Left column shows the results as a function of $\nu=\dc/\sigma_{cc}(M)$, i.e. in terms of the r.m.s. of the cold dark matter, while in the right column we set $\nu=\dc/\sigma_{mm}(M)$. It is evident that the description in terms of $\sigma_{mm}$ results in large departures from universality as $m_{\nu}$ is varied.  In contrast, using $\sigma_{cc}$ yields results which are much more universal.  

\begin{figure}[t!]
\begin{center}
\includegraphics[width=.98\textwidth]{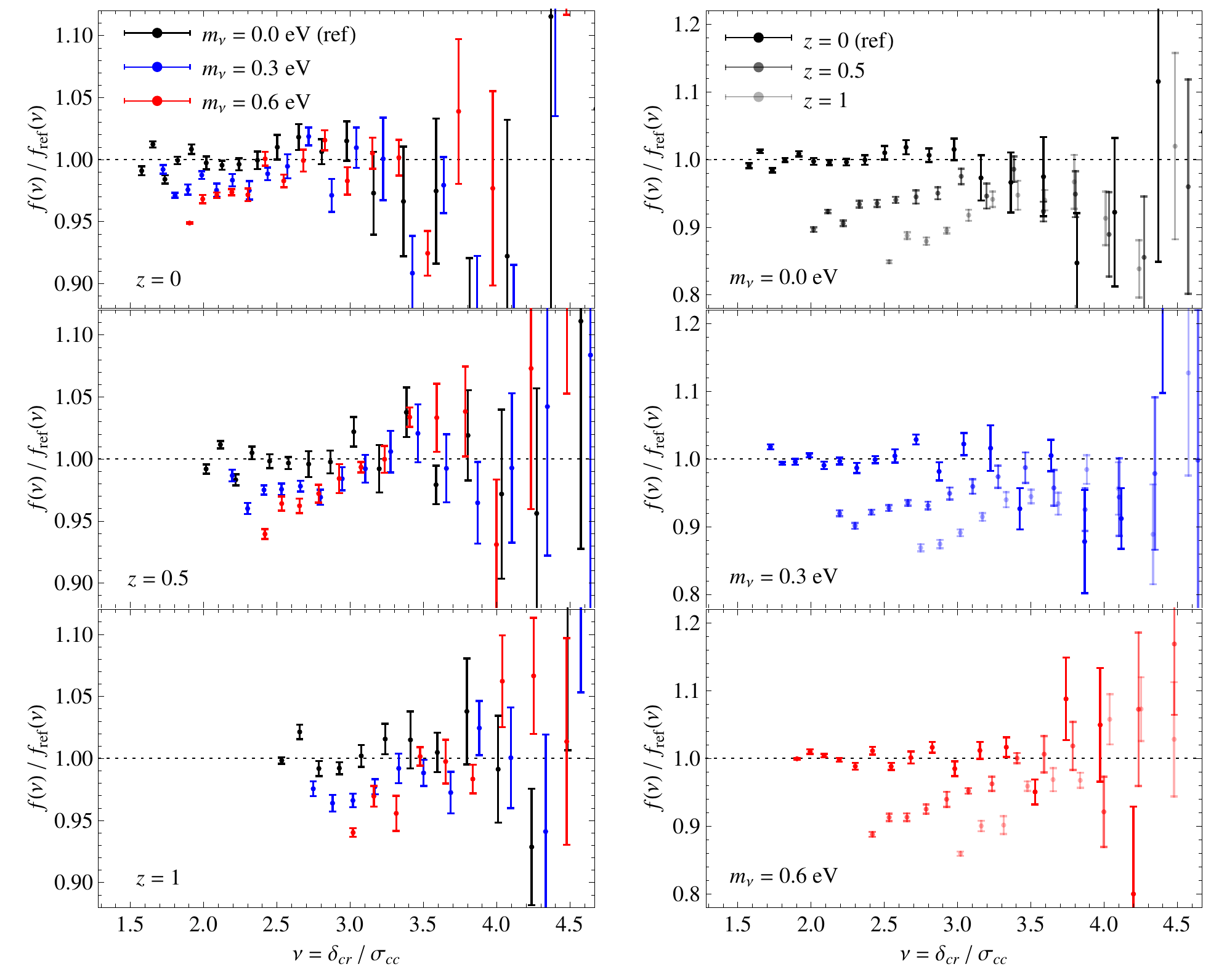}
\end{center}
\caption{\label{fig:mfnuzc} Left column: residuals of the measured $\nu f(\nu)$ w.r.t. to the best fit (assuming the ST form) to the $m_{\nu}=0$ data. Black, blue and red data points show the results respectively for the $m_{\nu}=0$, $0.3$ and $0.6$ eV models. Different panels correspond to $z=0$, $0.5$ and $1$, top to bottom. Right column: residuals of the measured mass function $f(\nu)$ w.r.t. to the best fit (assuming the ST form) to the $z=0$ data. Data points for the higher redshift outputs are shown by lighter shades of the given color. Different panels correspond to $m_{\nu}=0$, $0.3$ and $0.6$ eV, top to bottom.}
\end{figure}

Having confirmed our expectation that $\sigc$ is the relevant scaling variable, the analysis in the rest of this section will no longer consider $\sigm$. We notice that, even for $\sigc$ there is a small residual dependence on $m_\nu$ (left column of figure~\ref{fig:mfnu}). Figure~\ref{fig:mfnuzc} looks at this effect in more detail, by comparing it to the departure from universality as a function of redshift.  To emphasize the differences (which are small), the left column shows the ratio of the differential mass function to the {\em best fit} obtained for the ST functional form to the massless neutrino mass function for each given redshift ($z=0$, 0.5 and 1, top to bottom). The differences as a function of $m_\nu$ are $\nu$-dependent, but of the order of a few percent in the $m_{\nu}=0.6$ eV case, and slightly increasing with redshift. It should be stressed that the three neutrino cosmologies being compared share the same values for $\Omega_m$ and $A_s$. This results in values of $\sigec$ for the $m_{\nu}=0.3$ and $0.6$ eV cases that are smaller by 8 and 16\% respectively, w.r.t. the massless neutrino case.   

Departures from universality in redshift have already been reported and studied in the literature \cite{TinkerEtal2008, CrocceEtal2010}. These are shown on the right column of figure~\ref{fig:mfnuzc} where each panel shows, for each $m_\nu$, the ratio of the mass function $f(\nu)$ at $z=0$, 0.5 and 1 w.r.t. the best fit of the ST expression to the results at $z=0$. Increasing redshift is marked as a lighter shade of color. Discrepancies are, in this case as well, dependent on $\nu$ but of the order of 8-16\% respectively for $z=0.5$ and 1 at about $\nu=2.5$. These departures correspond to relative differences in $\sigma_{cc}$ of the order of 20-40\%, again for $z=0.5$ and 1, w.r.t. $z=0$.

We now look in more detail at the degeneracy between $m_{\nu}$ and $\sig_8$, the variance of matter fluctuations on a scale of $R= 8\Mpc$. One might expect that the effect of neutrino masses on the halo mass function can be re-absorbed by a suitable change of $\sigec$ in a standard $\Lam$CDM universe, in which case the only way to break the degeneracy being measurements at different redshifts, exploiting the differences in the linear growth factors. To explore this we use simulation set B. This allows us to compare an $m_{\nu}=0.6$ eV model (H6s8) with a massless neutrino model of set A (H0) that has the same value of $\sigma_{8,mm}=0.83$. In addition, we can compare two massless neutrino models (H0s8 and H0s8-CDM) to the $m_{\nu}=0.6$ eV model of set A (H6), matching the values of both $\sigma_{8,mm}$ and $\sigma_{8,cc}$ of the latter.  

\begin{figure}[t!]
\begin{center}
\includegraphics[width=0.98\textwidth]{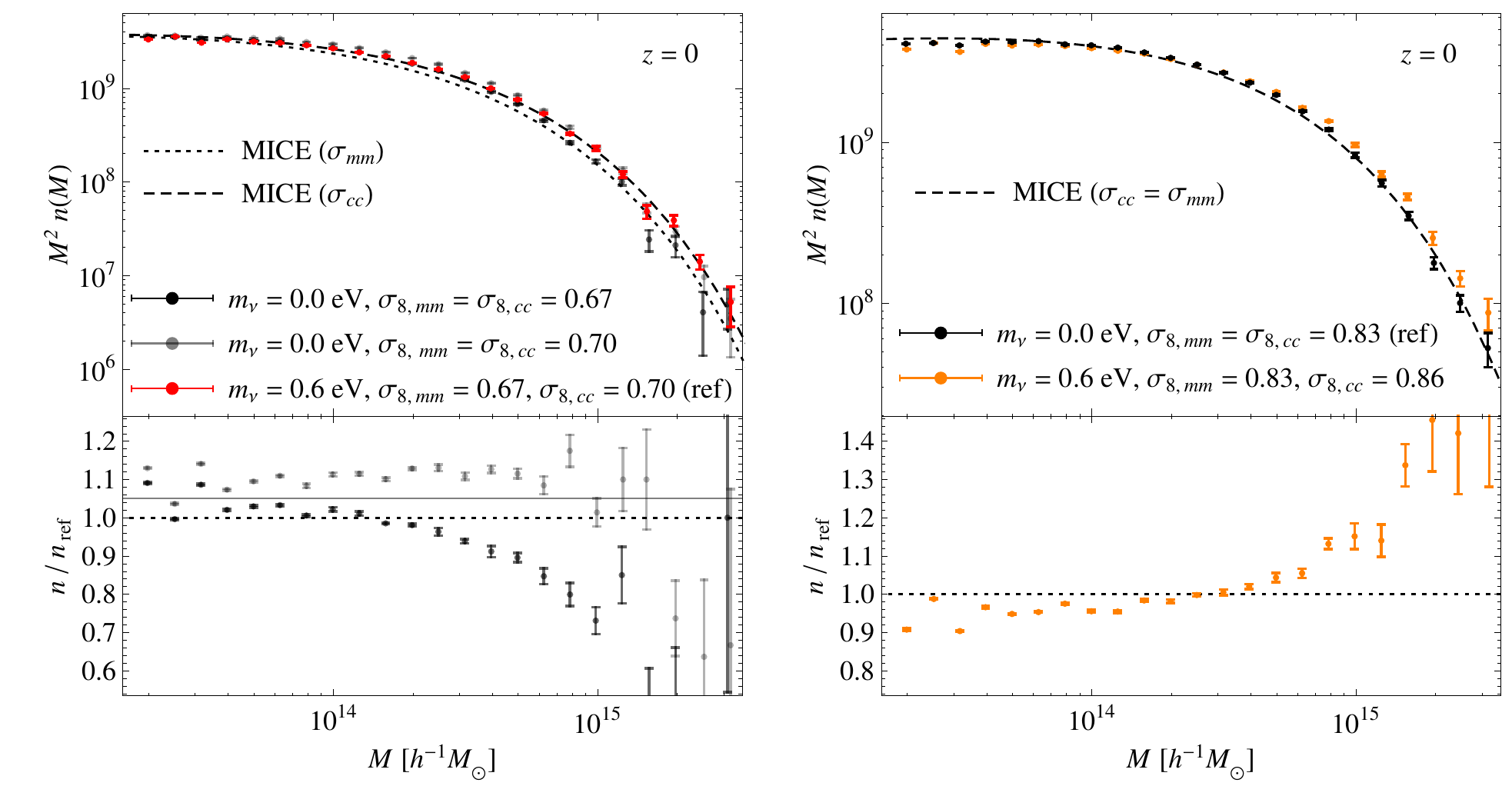}
\end{center}
\caption{\label{fig:mnm_s8} Degeneracy between $m_{\nu}$ and $\sigem$, $\sigec$: comparison of the mass functions as a function of the mass $M$ and at $z=0$ from $m_{\nu}=0.0$ eV and $m_{\nu}=0.6$ eV models, sharing the same value of $\sigec$ or $\sigem$. Left panels: comparison between model H6 (red data points) and models H0s8 (grey) and H0s8-CDM (light gray). The lower panel shows the ratio of H0s8 and H0s8-CDM to H6, with the black horizontal line showing the ratio $\Omega_c(\text{H0s8-CDM})/\Omega_c(\text{H6})$. Right panels: comparison between model H0 (black data points) and model H6s8 (orange). The lower panel shows the ratio of H6s8 to H0.}
\end{figure}

As we have seen, CDM halos are primarily sensitive to the relation between neutrino masses and the $\sig_8$ of the CDM species only. Therefore had we normalized the power spectrum of a $\Lam$CDM universe with the same $\sigem(z=0)$ of the correspondent neutrino cosmology, at fixed $\Omega_m$ we would have obtained different halo counts even at $z=0$, with larger discrepancies at higher redshifts. This is shown in figure~\ref{fig:mnm_s8}. Here, left panels present measurements of the halo mass function at $z=0$ as a function of mass for the models H6 in red, H0s8 in gray, and H0s8-CDM in light gray, where as expected from previous considerations the H6 cosmology has more objects than the H0s8 model. The difference with increasing $M$ increase as the ratio $\sigc(R)/\sigm(R)$ grows. The H0s8-CDM model lies much closer to the H6 model: the ratio of the two is almost constant and its difference from unity is mainly due to the different background density appearing on the r.h.s. of eq.~(\ref{eq:f}), shown as a continuous horizontal line in the residuals plot. The same arguments hold for the right panel of Figure~\ref{fig:mnm_s8}, where we compare the H0 run to H6s8, which has $m_{\nu}=0.6$ eV and $\sig_8$ matched to be the same as that of H0 at $z=0$. Again the halo counts differ substantially in the two simulations:  H6s8 has more halos because its CDM field has more power on all relevant scales than does the H0 run. 

\begin{figure}[t!]
\begin{center}
\includegraphics[width=.98\textwidth]{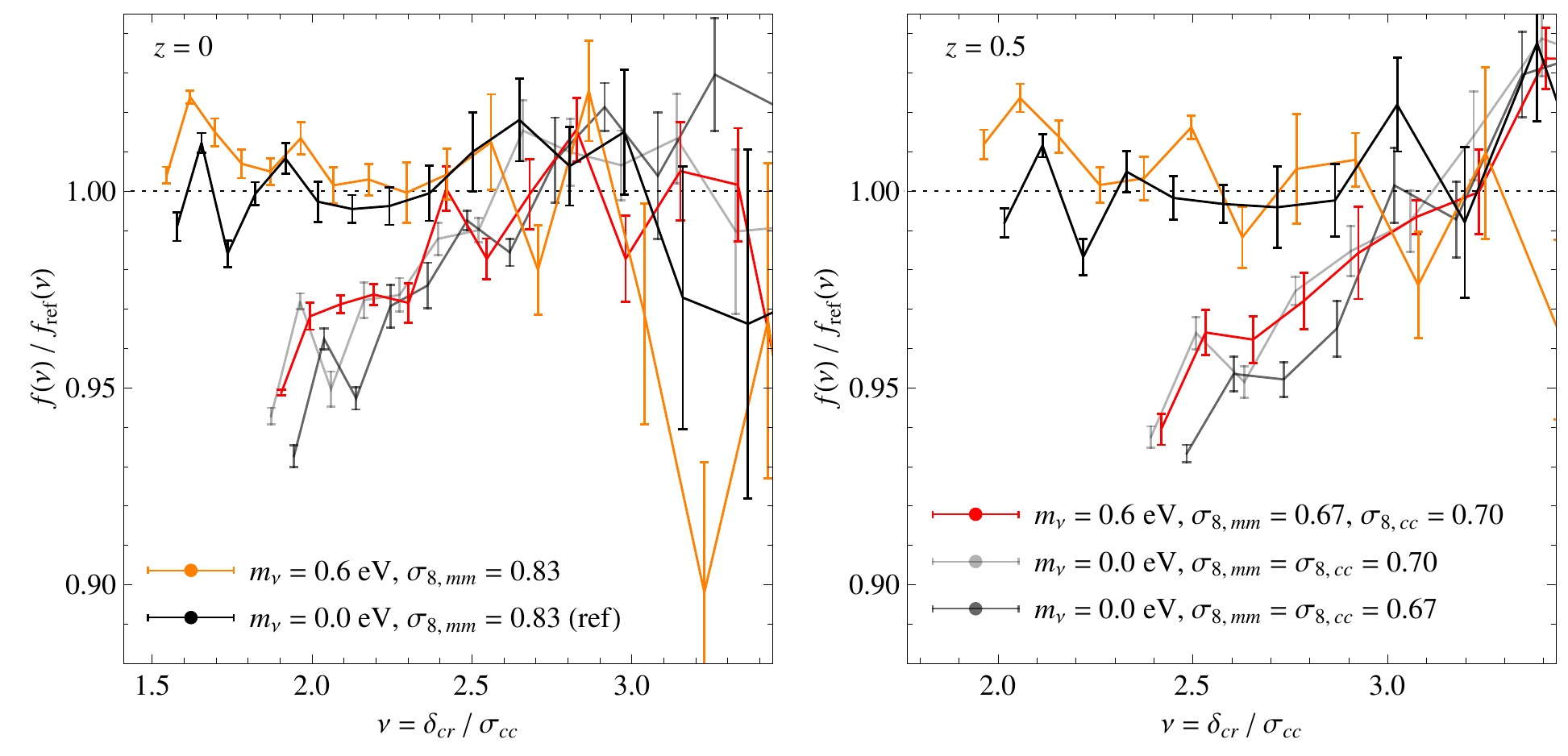}
\caption{\label{fig:mfnufits} Degeneracy between $m_{\nu}$ and $\sigem$, $\sigec$: ratio of the measured $f(\nu)$ for several models as a function of $\nu$ w.r.t. the best fit (assuming the ST functional form) to the massless neutrino model of set A, with $\sigem=\sigec=0.83$, taken as a reference (black data points). This is compared directly with a $m_\nu=0.6$ eV model with $\sigem=0.83$ and $\sigec=0.86$ (yellow points). In addition, the $m_\nu=0.6$ eV model with $\sigem=0.67$ and $\sigec=0.70$ from set A (red points) is compared with two massless neutrino models with both $\sigem=\sigec=0.67$ (gray points) and $\sigem=\sigec=0.70$ (lighter gray points). Left and right panels show the results at $z=0$ and $z=0.5$.}
\end{center}
\end{figure}

Again, we can learn something more about the small deviation of universality by looking at the same data in terms of the scaling variable $\nu=\dc / \sigc$. Figure~\ref{fig:mfnufits} compares all these models showing the ratio of the measured $f(\nu)$ as a function of $\nu$ w.r.t. the best fit (assuming the ST functional form) to the massless neutrino model of set A, taken as a reference. All models presented in figure~\ref{fig:mnm_s8} are now shown together, with the results at $z=0$ on the left panel, and those at $z=0.5$ on the right.  We can see that models even with very different $m_\nu$ have similar $\nu f(\nu)$ mass function provided that they have similar $\sigma_8$. Although error bars do not allow a clear indication, this agreement is even better if $\sigec$ rather than $\sigem$ is matched.  

Small deviations from universality, such as those we see here, are expected in Peaks/Excursion Set theory \cite{ParanjapeShethDesjacques2013} where cosmologies having the same value of $\sigec$ aare expected to have similar $n(M)$, but there would still be small differences because the predicted halo mass function depends on the slope of the power spectrum where $\sigc=\dc$ as well, and this we have not matched.

The main result of this section is that the proper variable to describe the halo mass function in a neutrino cosmology is the linear CDM power spectrum $P_{cc}(k)$, since only when this is done does $\de n / \de M$ behave as universally as in standard neutrino-less $\Lam$CDM universes.  E.g., only if one uses $\nu = \dc / \sigc$ is the violation of universality of order $10\%$ or smaller at $z=1$ (see figure~\ref{fig:mfnuzc}), compatible with findings from other groups \cite{TinkerEtal2008, CrocceEtal2010, WatsonEtal2013}.  We also reported small deviations from universality with respect to the different cosmologies in our set of models, at a few \% level, that have to be further investigated with reduced error bars. The overall picture is enforced by the simulation set B that we understand, to a large extent, again in terms of a description based on CDM perturbations only.

We believe these results will have important implications for cosmological parameters inference from galaxy clusters observations. In all previous cosmological analyses \cite{HasselfieldEtal2013, ReichardtEtal2013, Planck2013SZ} the total matter fluctuations, i.e. $\sig=\sigm$, have been used for the prediction of the mass function, leading to possible systematic errors in the derived constraints. The difference is not limited to the expected number of galaxy clusters, which
are more numerous if one uses the cold dark matter matter spectrum, but, most importantly, it affects the universality of the halo mass function with respect to cosmology. A key assumption in 
cosmological analyses is that the shape of the mass function is insensitive to changes in the
background cosmology when the total matter power spectrum is used, while we have shown that this
is not the case for massive neutrino models. For cosmological analyses this effect is even more relevant than the relative difference in the same cosmology of the expected number counts from the two prescriptions for the power spectrum. An estimate of such systematics, in a bayesian analysis of current clusters data, is the subject of our Paper III \cite{CostanziEtal2013B}.

\section{Halo bias}
\label{sec:bias}

In $\Lam$CDM models with massless neutrinos the relation between halo overdensity $\delta_h$ and the matter overdensity $\delta_m$ on very large scales is expected to be linear and deterministic, i.e. (see, e.g. \cite{BernardeauEtal2002})
\be
 \delta_h(x) = b\,\delta_m(x)\,,
\ee
with a constant bias parameter $b$, resulting in the simple expression for the halo power spectrum given by $P_{hh}(k) = b^2\,P_{mm}(k)$. Additional nonlinear but local corrections in the equation above induce a scale-dependence in the relation between halo and matter power spectra and are however necessary to describe higher-order correlations. 

In a cosmology with massive neutrinos defining halo bias in terms of the total or cold matter density field is, in principle, a matter of convenience. Nevertheless given the scale-dependent difference between $P_{mm}$ and $P_{cc}$ we can expect an additional scale-dependence in $P_{hm}$ or $P_{hc}$, relevant at relatively large-scales, resulting simply from a ``wrong'' choice. In other words, if bias is constant on large-scales in one case, it cannot
be so in the other.  We will show that -- not surprisingly after the
results of the previous section -- formulating the problem is in terms of
the cold dark matter field is the right thing to do.

To proceed, we define auto- and cross- bias factors with respect to the CDM-field:  
\bea
\label{eq:biasc}
 b_c^{(hh)} & \equiv & \sqrt{\frac{P_{hh}(k)}{P_{cc}(k)}}\,,\\
 \label{eq:biasxc}
 b_c^{(hc)} & \equiv & \frac{P_{hc}(k)}{P_{cc}(k)}\,,
\eea   
as well as the analogous quantities 
\bea
\label{eq:biasm}
 b_m^{(hh)} & \equiv & \sqrt{\frac{P_{hh}(k)}{P_{mm}(k)}}\,,\\
 \label{eq:biasxm}
 b_m^{(hm)} & \equiv & \frac{P_{hm}(k)}{P_{mm}(k)}\,,
\eea   
for the bias with respect to $P_{mm}$.  Previous work in neutrino-less cosmologies has shown that the bias factors from the cross-correlations tend to be a few percent larger than those from auto-correlations \cite{ManeraShethScoccimarro2010, PollackSmithPorciani2013}.

\begin{figure}[t!]
\begin{center}
\includegraphics[width=.98\textwidth]{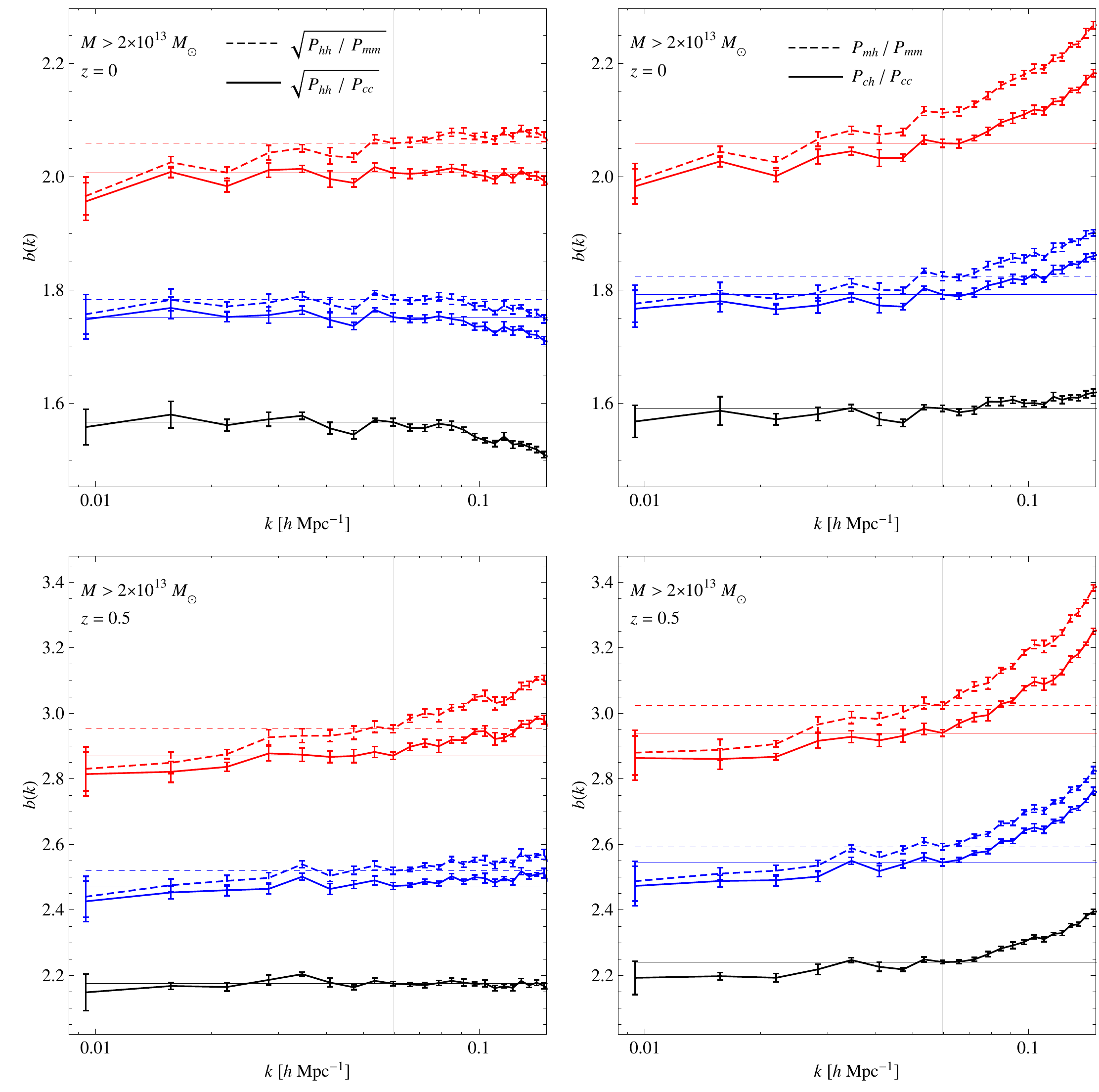}
\caption{\label{fig:biasM1} 
Halo bias as a function of scale determined from the simulation Set A for halos with $M>2\times10^{13}\Ms$. Left panels show the measurements of linear bias $b_c^{(hh)}$ ({\em continuous curves}) and $b_m^{(hh)}$ ({\em dashed curves}) from the halo power spectrum $P_{hh}(k)$. Right panels show $b_c^{(hc)}$ ({\em continuous curves}) and $b_m^{(hm)}$ ({\em dashed curves}) respectively from the $P_{hc}$ and $P_{hm}$ cross-power spectra. Top left panels correspond to $z=0$, bottom panels to $z=0.5$. The continuous and dotted horizontal lines show the constant bias values determined from measurements of $b_c$ and $b_m$, respectively, at $k=0.07\kMpc$, shown in turn as a vertical gray line in all panels. Error bars show the uncertainty on mean over the eight realizations.}
\end{center}
\end{figure}

\begin{figure}[t!]
\begin{center}
\includegraphics[width=.98\textwidth]{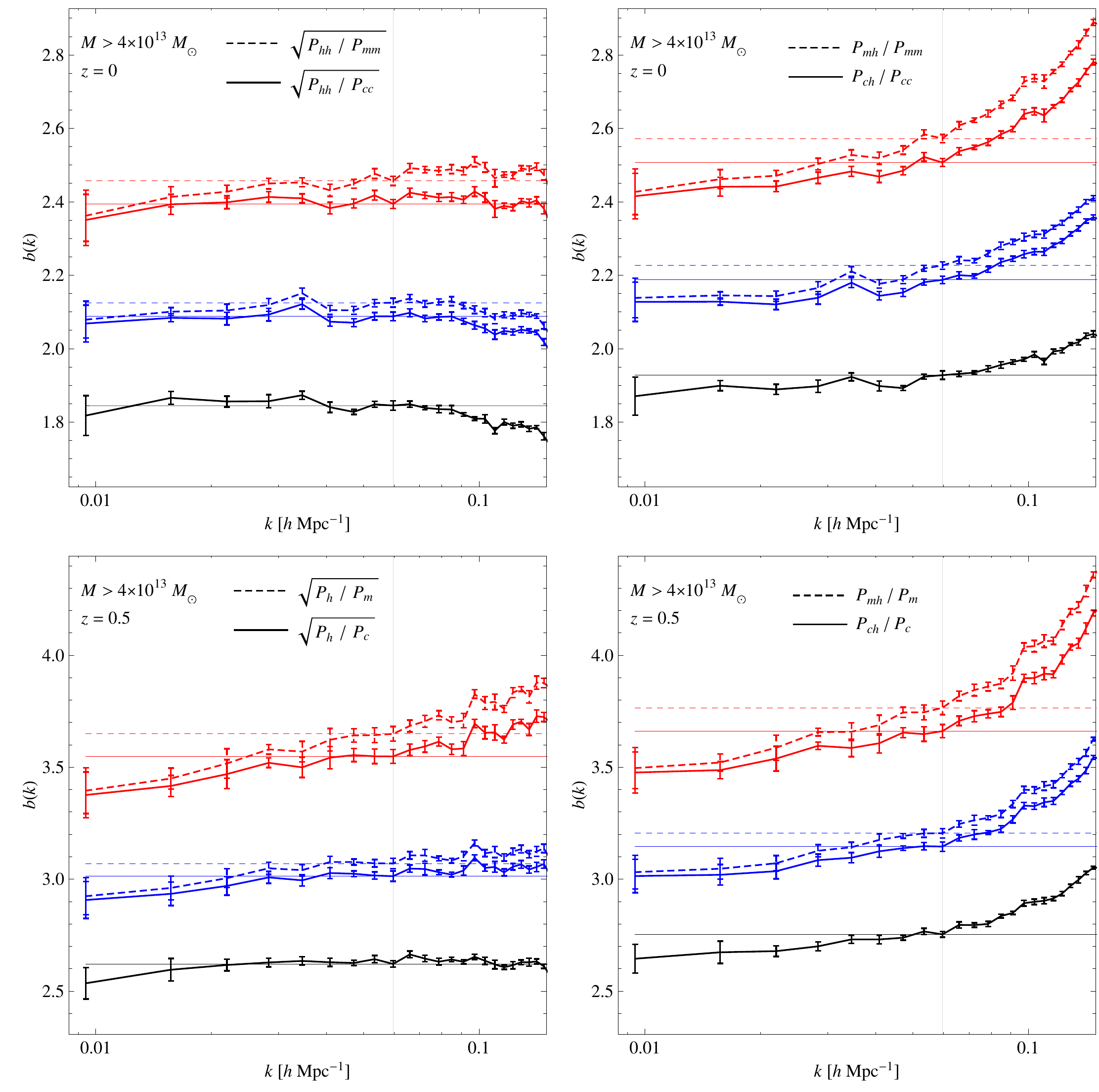}
\caption{\label{fig:biasM2}
Same as previous figure~\ref{fig:biasM1} but for halos with $M>4\times10^{13} \Ms$.}
\end{center}
\end{figure}

For each of the simulation sets in Table~\ref{tab:par} we measured these bias factors for two halo populations: one has $M>2\times 10^{13}\Ms$ and the other $M>4\times 10^{13}\Ms$, irrespective of redshift and cosmological parameters.  

\begin{figure}[t]
\begin{center}
\includegraphics[width=.98\textwidth]{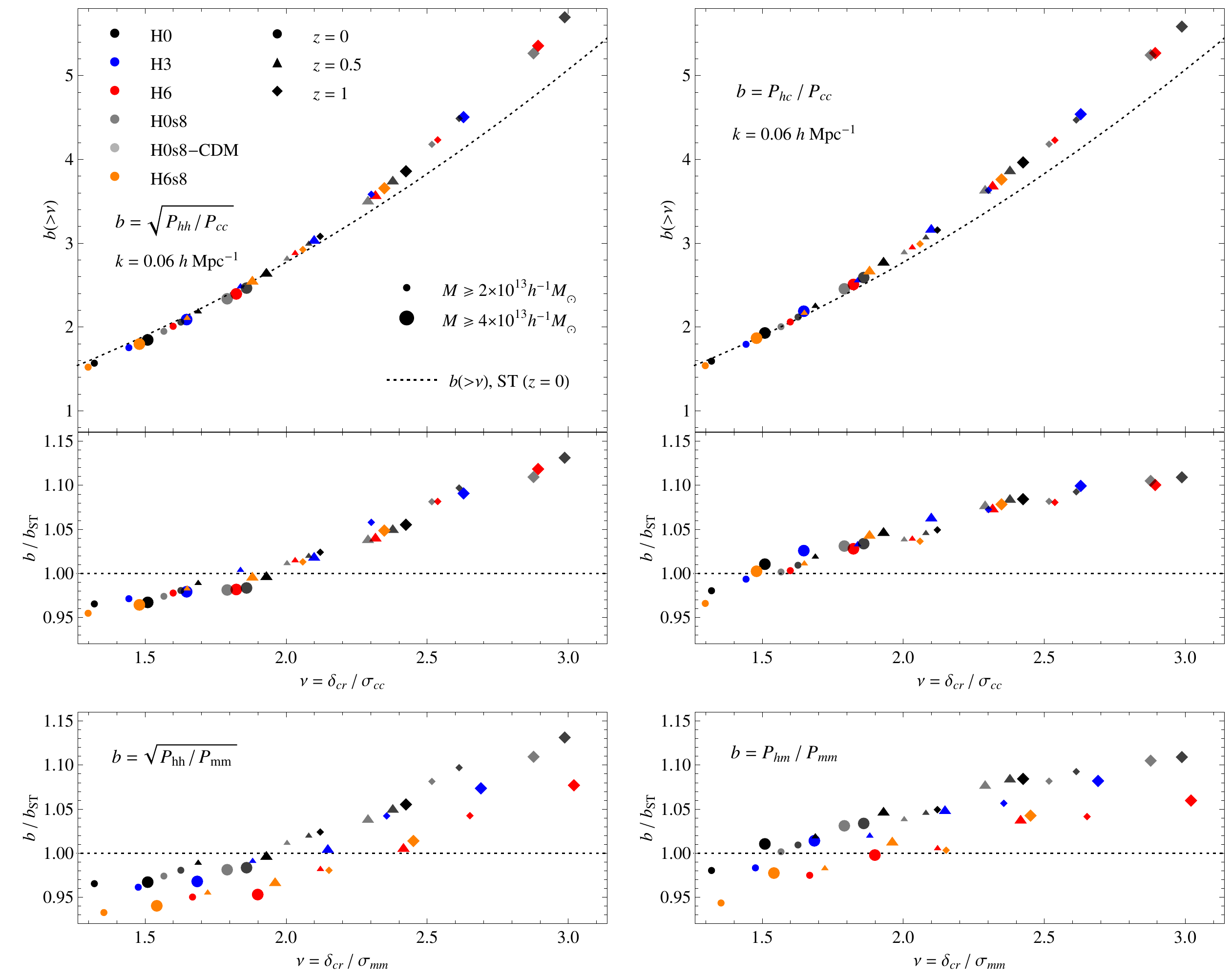}
\caption{\label{fig:biasnu}
Measurements of the linear bias coefficient as a function of the value of $\nu=\dc/\sigma$ corresponding to the mass threshold from the halo power spectrum ({\em left panels}) and halo-matter cross-power spectrum ({\em right panels}) at three different  redshifts ($z=0$, 0.5, 1), for two mass thresholds ($M>2$ and $4\times 10^{13}\Ms$) and for all the cosmologies under consideration. Theoretical predictions for the linear bias from the standard ST mass function evaluated for the H0 model at $z=0$ are shown by the dotted curve. The middle panel shows the residuals w.r.t. the ST prediction. Top and middle panels assume halos to be biased w.r.t. the cold DM perturbations and therefore assume $\nu=\dc/\sigc$. The bottom panels shows the residuals w.r.t. ST assuming instead halos to be biased w.r.t. the total matter perturbations and $\nu=\dc/\sigm$.
}\end{center}
\end{figure}

We focus first on a comparison between the different halo bias definitions described above, $b_m$ and $b_c$. Figure~\ref{fig:biasM1} shows the bias, as a function of scale, determined for halos with $M>2\times10^{13} \Ms$ while figure~\ref{fig:biasM2} shows the same measurements for halo populations determined by $M>4\times10^{13} \Ms$. Left column shows $b_c^{(hh)}(k)$ (symbols with connecting continuous curves) and $b_m^{(hh)}(k)$ (symbols with dashed curves) defined respectively in eqs.~(\ref{eq:biasc}) and~(\ref{eq:biasm}) for the three models of Set A. The right column shows $b_c^{(hc)}(k)$ (symbols with connecting continuous curves) and $b_m^{(hm)}(k)$ (symbols with dashed curves) defined in eqs.~(\ref{eq:biasxc}) and (\ref{eq:biasxm}) in terms of cross-power spectra. Top row shows the results at $z=0$, bottom row at $z=0.5$. To guide the eye, the continuous and dashed horizontal lines show the values of $b_{c}(k)$ and $b_{m}(k)$ at $k=0.07\kMpc$ (shown as a vertical line):  below this value of $k$, the bias is observed to be constant for most measurements.  This value also defines the bias values we use for the study of bias as a function of $\nu$ later in figure~\ref{fig:biasnu}. 

The fixed mass threshold clearly results in different bias values for the three models.  This is a consequence of the fact that the same mass threshold corresponds to quite different number counts for the different models \cite{ShethTormen1999}.  The figure also illustrates the different scale-dependence of the two bias definitions. The departure from a constant value is more evident for the bias defined w.r.t. $P_{mm}$. This is consistent with eq.~(\ref{eq:biasc}) and with the analysis of $n(M)$ presented in section~\ref{sec:massf}, reinforcing our understanding of observables in massive neutrino cosmologies in terms of the CDM distribution. However, a noticeable $k$-dependence is also present for the $b_c$ measurements, increasing with the value of the bias itself. This might be due to nonlinear effects that we ignore in our comparison and are the subject of ongoing work. 

Although the auto- and cross- values differ slightly, both $b_c$ and $b_m$ converge to the same values on the largest scales (smallest $k$ values) probed by the simulations, reflecting the fact that for $k\lesssim k_{nr}$ the DM and CDM linear power spectra are the same. For $k>k_{nr}$ but still in the linear regime, both $b_m^{hh}$ and $b_m^{hm}$ exhibit scale dependence, but have the same asymptotic behavior, given that
\be
 b_{m}^{(hh)} \equiv \sqrt{\frac{P_{hh}}{P_{mm}}}= b_{c}^{(hh)} \sqrt{\frac{P_{cc}}{P_{mm}}}\;\xrightarrow{ k\gg k_{nr}} b_{c}^{(hh)} (1-f_{\nu})\,,
\ee
and
\be
 b_m^{(hm)} \equiv \frac{P_{hm}}{P_{mm}}= b_c^{(hc)} \frac{P_{cm}} {P_{cc}}\;\xrightarrow{ k\gg k_{nr}} b_c^{(hc)}(1-f_{\nu}) \,.
\ee
At $k=0.07\kMpc$, the $b_{m}^{(hh)}$ coefficients are 5\% larger than the corresponding $b_{c}^{(hh)}$ values for $m_{\nu}=0.6$ eV.

\begin{figure}[t]
\begin{center}
\includegraphics[width=0.98\textwidth]{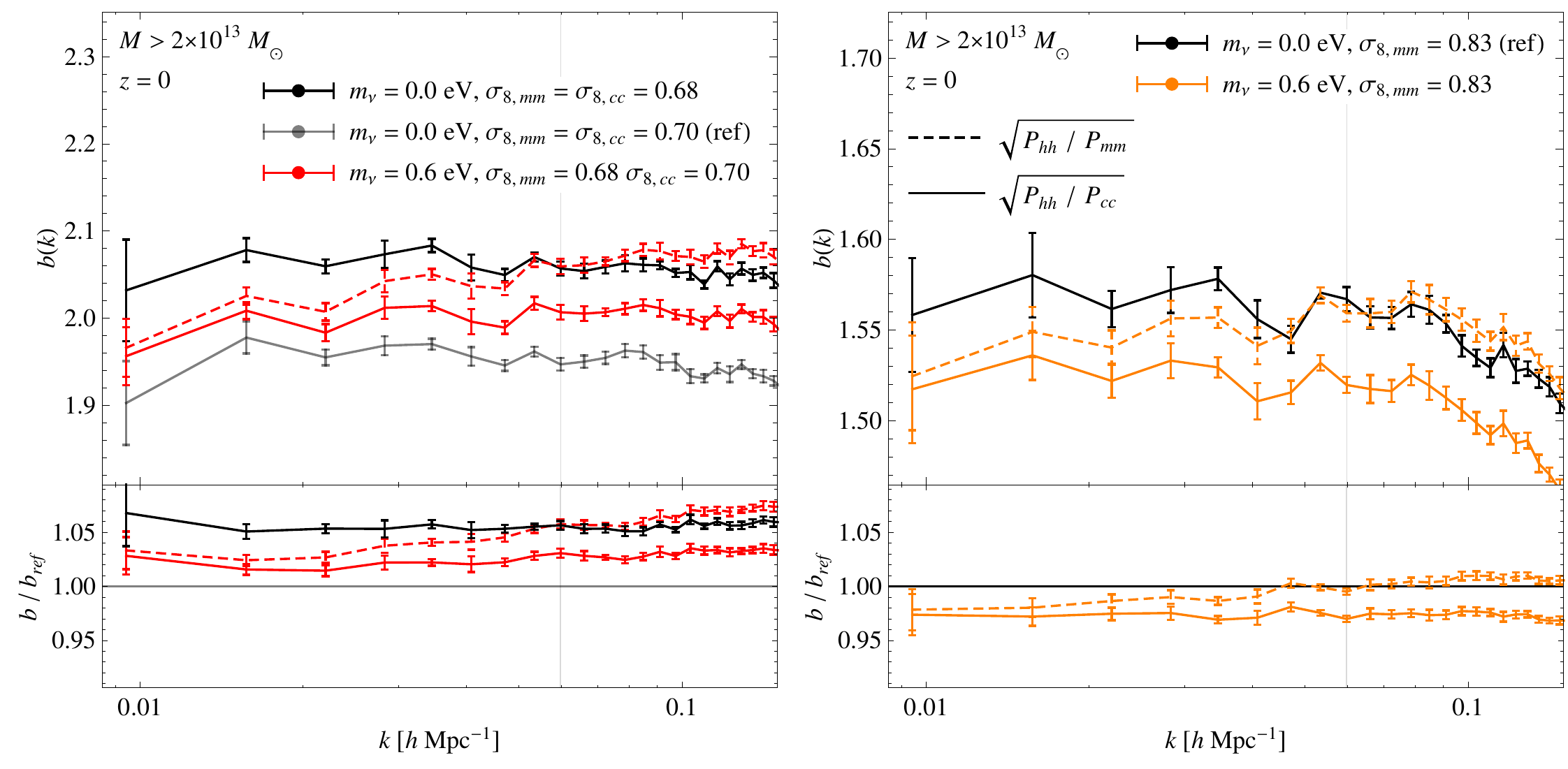}
\caption{\label{fig:bias_s8}
Degeneracy $m_{\nu}$-$\sig_8$: halo bias at $z=0$ determined from the simulation set B. Left panel: comparison of the H6 model (in red) to the H0s8 (dark gray) and H0s8-CDM (light gray) models. As before continuous lines correspond to bias with respect to the cold dark matter power spectrum, while dashed lines to bias with respect to the total matter power spectrum. Right panel: comparison of the H0 model( in black) to the H6s8 model (in yellow), the two sharing the same value of $\sigem$. In both panels we considered only halos with $M>2\times10^{13} \Ms$.}
\end{center}
\end{figure}

It is interesting to compare now the bias values in models with and without massive neutrinos that have the same value of $\sigma_8$.  The left panel of figure~\ref{fig:bias_s8} compares the results for the lowest mass threshold at $z=0$ of the model with $m_{\nu}=0.6$ eV (H6, red data points), already shown in figure~\ref{fig:biasM2}, with those of the two massless neutrino models from simulation Set B, matching in turn the value of $\sigem$ (H0s8, black data points) and $\sigec$ (H0s8-CDM, gray data points), the latter taken as reference model. 

 At fixed mass threshold objects in the H0s8 runs are more clustered than in the H6 and H0s8-CDM runs, as expected from measurements of the mass function (see figure~\ref{fig:mnm_s8}). Despite these marginal differences, however, it is important to notice that the three models are characterized by a bias $b_{c}(k)$ with the same dependence on scale. This is particularly evident in terms of their relative differences, shown in the lower panels, which are identical for both $b_c^{(hh)}$ and $b_c^{(hc)}$. One could argue that the agreement between the bias measured in such different models could be further improved by appropriately rescaling the spectral index of the linear power spectrum of the model H0s8-CDM to match the one of H6 at the scale where $\sigma_{cc}=\delta_{cr}$.  Doing so would not fix the additional scale-dependence of $b_m(k)$.     

Similar conclusions can be drawn from the right panel of figure~\ref{fig:bias_s8} where we show, instead, a comparison of the H0 model to the H6s8 model, sharing the same value of $\sigem$.

In recent analyses which constrain neutrino masses using galaxy clustering in large redshift surveys  bias is treated as a free parameter over which to marginalize (see, e.g. \cite{ZhaoEtal2013, GiusarmaEtal2013}).  The standard practice defines the bias with respect to the total matter power spectrum, and assumes that $b_m$ is scale-independent up to some $k_{max}$.  Our analysis shows that bias coefficients defined in this way are, in fact, scale-dependent, even at linear scales. Therefore, to avoid systematics inaccuracies, future analysis of galaxy surveys data must account for this.  

The final part of this section is devoted to study the universality of bias, measured at a fixed scale $k$ and described as a function of the variable $\nu=\dc/\sigma$ as the mass threshold, the redshift and cosmological models are varied. If $\nu f(\nu)$ is universal, then 
\be\label{eq:bE}
 b = 1 - \frac1{\dc}\frac{\de\ln \, f(\nu)}{\de\ln\nu} \,,
 \ee
is also a universal function of $\nu$.  However, our measurements correspond to 
\be
\label{eq:bMean}
 \bar{b}(>M_{min}) = \frac{\int_{M_{min}} b(M)\, n(M)\, \de M}{\int_{M_{min}} n(M)\, \de M}\,,
\ee
with $M_{min} = 2\times 10^{13}\Ms$ and $4\times 10^{13}\Ms$; in terms of the scaling variable, this reads 
\be
 \label{eq:bMeannu}
 \bar{b}(>\nu_{min}) = \frac{\int_{\nu_{min}} b(\nu)\,[f(\nu)/M(\nu)]\,\de\nu}{\int_{\nu_{min}} [f(\nu)/M(\nu)]\,\de\nu}\,,
\ee
where $\nu_{min}=\dc/\sigma(M_{min})$ and where the presence of the factor $1/M$ in the integrand does not allow $\bar{b}$ to be a function of $\nu_{min}$ alone.  In principle, this introduces an explicit dependence on cosmology, even if $\nu f(\nu)$ and $b(\nu)$ themselves are universal. 

Figure~\ref{fig:biasnu} shows the measurements of the linear bias {\em at} $k=0.06\kMpc$ from both the halo power spectrum (left panels) and halo-matter cross-power spectrum (right panels) at three different  redshifts ($z=0$, 0.5 and 1), for two mass thresholds ($M>2$ and $4\times 10^{13}\Ms$) and for all the neutrino cosmologies (Set A and B), as a function of the value of $\nu=\dc/\sigc$. The measured values are compared, in the upper panels, to the prediction for the standard ST mass function and bias (dotted curves) computed from eq.~(\ref{eq:bMean}) for the massless neutrino cosmology at redshift zero. Due to the small residual scale-dependence, a different choice of the value of $k$ leads to the same results, with more noise at large scales. 

We notice that the dependence of such predictions on cosmology and redshift are very small. Indeed, this fact can be appreciated from the data points themselves: all results from different masses, redshifts and cosmologies are aligned as one would expect for a function of $\nu$ alone. Small departures from such overall behavior can be seen in the middle panels, showing the residuals w.r.t. the ST predictions. Notice that we do expect some departure from universality both from the mass function results as from the definition of $\bar{b}(>M)$ itself. These effects are, nevertheless, much smaller than those obtained defining bias from the total matter power spectrum and plotting the results as a function of $\nu=\dc/\sigm$, as shown in the lower panels of figure~\ref{fig:biasnu}, separated for clarity. Here, we show only the ratio w.r.t. the ST prediction, the latter coinciding with the one used in the middle panels to allow a direct comparison. 
 
We have seen that, if we adopt as a definition for the peak height $\nu=\dc/\sigc$, linear bias is to a very good approximation a universal function with respect to redshift even if the halo mass function is not, in agreement with previous findings \cite{TinkerEtal2010}. This can be understood from the PBS argument, where bias coefficients are defined as a logarithmic derivative of the dimensionless mass function $f(\nu)$, eq.~(\ref{eq:bE}). Linear bias is, to a large extent, a universal function, since most significant departures from universality w.r.t. redshift in the mass function come as changes in the amplitude, independently of the halo mass.


\section{Conclusions}
\label{sec:conclusions}

We have investigated the effects of massive neutrinos on halo abundance and clustering taking advantage of a large set of numerical N-body simulations, implementing neutrinos as particles. We have shown evidence, for the first time, that the halo mass function for these models can be reproduced by existing fitting formulae derived from simulations of standard $\Lam$CDM cosmologies  only if evaluated in terms of the variance of small-scale cold dark matter perturbations. This description has been proposed, as a result of theoretical investigations, in \cite{IchikiTakada2012}. Previous studies commonly assumed, however, a dependence on the total matter density variance, leading to large and ultimately artificial departures from universality as the total neutrino mass is varied. 
   
Our results are based on halo catalogs determined by means of a FoF halo finder accounting for CDM particles only. Lacking a proper definition of a FoF finder for neutrino particles we tested our results by considering as well catalogs obtained including all particles, finding percent-level differences in the mass function for low masses. Such differences are expected to be even smaller once unbound neutrino particles with large thermal velocities are removed from the halo by a suitable algorithm. The emerging picture shows neutrinos with masses in the currently viable range playing a minor role in the nonlinear collapse of cold DM structures: even though a fraction of them do cluster inside CDM halos, their contribution to the total mass of a halo is negligible \cite{SinghMa2003, RingwaldWong2004, BrandbygeEtal2010, VillaescusaNavarroEtal2011, Villaescusa-NavarroEtal2013}.

We studied in detail the halo abundance as a function of the variable $\nu = \dc / \sig$, showing that while universality is recovered to a large extent by setting $\sigma=\sigc$, minor departures with respect to neutrino masses are detectable and comparable to those already seen in standard $\Lam$CDM models for instance as a function of redshift \cite{TinkerEtal2008, CrocceEtal2010}. This implies that neutrino masses are primarily degenerate with the amplitude of the fluctuactions in the CDM field $\sigec$, and, to a lesser extent, to the slope of the spectral index on the scale where $\sigec=\dc$.

This implies that neutrino masses are primarily degenerate with the amplitude of the fluctuactions in the CDM field, $\sigec$. 

The impact of these findings on cosmological analyses from galaxy clusters abundances can be significant. An estimate of the systematic error induced by the wrong assumption for the mass function dependence on the linear power spectrum and an application to recent data-sets is the subject of our companion paper, \cite{CostanziEtal2013B}.

In the second part of this work we studied halo clustering, identifying a definition of halo bias in terms of cold dark matter perturbations as the only one able to recover the expected constant bias parameters at large scales. This is, to a large extent, a natural consequence of the results on the mass function. We notice some small, residual scale-dependence for large bias values possibly due to nonlinear effects. However, bias coefficients computed from the total dark matter power spectrum are significantly scale-dependent. A comparison of bias measurements as a function of $\nu = \dc / \sigc$ shows remarkable universality, in stark contrast with the total matter description with variable $\nu = \dc / \sigm$.

Our results on bias have, as well, important implications for the analysis of galaxy clustering aiming at constraining neutrino masses. In previous works, bias parameters defined in terms of total matter perturbations are assumed as constant and marginalized over, introducing a non-negligible systematic error in the results. An estimate of the effect of these assumptions, as well as a detailed study of nonlinear bias, including bispectrum measurements, will be the subject of future work. 

\bigskip

\section*{Acknowledgements} 
We are grateful to Julien Lesgourgues for useful discussions. Calculations were performed on SOM2 and SOM3 at IFIC and on the COSMOS Consortium supercomputer within the DiRAC Facility jointly funded by STFC, the Large Facilities Capital Fund of BIS and the University of Cambridge, as well as the Darwin Supercomputer of the University of Cambridge High Performance Computing Service (http://www.hpc.cam.ac.uk/), provided by Dell Inc. using Strategic Research Infrastructure Funding from the Higher Education Funding Council for England.  FVN and MV are supported by the ERC Starting Grant ``cosmoIGM''. MV is also supported by I.S. INFN/PD51. ES and RS were supported in part by NSF-AST 0908241.

\bibliographystyle{JHEPb}
\bibliography{}

\end{document}